\documentclass[12pt]{article}
\usepackage{bm}
\usepackage[final]{ametsoc}
\usepackage{amsmath}
\usepackage{epsfig}

\begin{document}







\title{\bf Stratocumulus over SouthEast Pacific: Idealized 2D simulations with the Lagrangian Cloud Model\\}
\author{{M. Andrejczuk, A. Gadian, A. Blyth} \\
\vspace{-0.2in}
Corresponding Author: \\
\vspace{-0.2in}
M. Andrejczuk \\
M.Andrejczuk@leeds.ac.uk \\
\vspace{-0.2in}
School of Earth and Environment \\
\vspace{-0.2in}
University of Leeds \\
\vspace{-0.2in}
Leeds, LS2 9JT \\
\today }

\amstitle

\pagenumbering{arabic}

\begin{abstract}
In this paper a LES model with Lagrangian representation of microphysics 
is used to simulate stratucumulus clouds in idealized 2D set-up based on 
the VOCALS observations. The general features of the cloud simulated by the 
model, such as cloud water mixing ratio and cloud droplet number profile agree well 
with the observations.  
The model can capture observed relation between aerosol 
distribution and concentration measured below the cloud and cloud droplet number. Averaged over the 
whole cloud droplet spectrum from the numerical model and observed droplet 
spectrum are similar, with the observations showing a higher concentration of 
droplets bigger than 25 ${\mu}$m.  Much bigger differences are present when 
comparing modelled and observed droplet spectrum on specific model level. 

Despite the fact that microphysics is formulated in a Lagrangian framework 
the standard deviation of the cloud droplet distribution is larger than 1 ${\mu}$m. 
There is no significant narrowing of the cloud droplet distribution in the up-drafts, but the distribution 
in the up-drafts is narrower than in the down-drafts. Modelled and observed standard 
deviation profiles agree well with observations for moderate/high cloud droplet 
numbers, with much narrower than observed droplet spectrum for low droplet number. 

Model results show that a significant percentage of droplets 
containing aerosol bigger than 0.3 ${\mu}$m didn't reach activation radius, 
yet exceeding 1 ${\mu}m$, what is typically measured as a cloud droplets. 
Also, the relationship between aerosol sizes and cloud droplet sizes is complex; 
there is a broad range of possible cloud droplet sizes for a given aerosol size. 
\end{abstract}
\newpage
\noindent {\bf 1. Introduction}
Microphysics parametrization is still an unresolved problem in numerical models. 
The large number of aerosol particles and cloud droplets in the atmosphere force 
the use of the simplified description of this process. The simplest approaches, 
single moment bulk models (\citeauthor{kessler},~\citeyear{kessler}), 
assume that the cloud is in 
equilibrium with the air. 
Water condenses/evaporates within one time-step and the final state is determined by 
the thermodynamic conditions. These type of models evolved to multi-moment models 
predicting not only the mass of the water, but also the number of droplets 
(\citeauthor{Koenig}~\citeyear{Koenig}; \citeauthor{Ferrier}~\citeyear{Ferrier};
\citeauthor{Meyers}~\citeyear{Meyers}; \citeauthor{Khairoutdinov}~\citeyear{Khairoutdinov};
\citeauthor{Seifert}~\citeyear{Seifert}; \citeauthor{Milbrandt}~\citeyear{Milbrandt};
\citeauthor{Morrison}~\citeyear{Morrison}).  
In the bulk approach, since the  
droplet spectrum is not predicted   
there is a problem with determining the amount of the water transported from the cloud by 
precipitation and how fast this transport proceeds. Although bulk models provide
 some useful information their applications, especially, when precipitation is present 
or information about spectrum is important is limited. Such models, however, are widely used 
due to their simplicity and fast execution. 

Spectrum predicting models, bin models, replace the discrete description of individual cloud 
droplets with its continuous counterpart and are used in numerical models since 1970's. 
There are variety of such models (~\citeauthor{flossmann}~\citeyear{flossmann};
\citeauthor{ackerman1}~\citeyear{ackerman1};
\citeauthor{khain2004}~\citeyear{khain2004};
\citeauthor{kogan1999}~\citeyear{kogan1999}) differing in the way droplet activation/deactivation 
is handled and how condensational growth and collision operator is calculated. Such 
models provide detail information about the structure of the cloud and cloud droplet 
spectra and were successfully compared with the aircraft observations 
(\citeauthor{flossmann}~\citeyear{flossmann};\citeauthor{kogan1999},~\citeyear{kogan1999}). 
These models are used not only in a Large Eddy Simulation framework, 
but also in much larger scales, for instance to investigate effect of aerosol on hurricanes 
(\citeauthor{khain2009},~\citeyear{khain2009}), as well as 
much smaller scales to investigate details of the cloud/clear air mixing processes 
(\citeauthor{andrejczuk2004},~\citeyear{andrejczuk2004}).
However, there are still unresolved issues 
in the bin approach, like how efficiently include information about aerosol, 
activation and/or deactivation of the aerosol or artificial 
spectrum broadening during condensational growth. Some of these issues 
are not present for 2 dimensional bin models, 
\citeauthor{clark1973} ~(\citeyear{clark1973}),
\citeauthor{ovchinnikov2010} ~(\citeyear{ovchinnikov2010}) 
which represent both aerosol and 
cloud droplet size, but such models are not in common use.

There are alternative numerical model approaches to the continuous droplet spectrum 
representation. Two models using Lagrangian approach to 
droplet/aerosol spectrum representation were developed recently - Lagrangian 
Cloud Model by \citeauthor{andrejczuk2008}~(\citeyear{andrejczuk2008}), 
later extended to include the collision-coalescence process 
\citeauthor{andrejczuk2010}~(\citeyear{andrejczuk2010}) and 
a super-droplet approach by \citeauthor{shima}~(\citeyear{shima}). 
In this new formulation, Lagrangian parcels representing millions of physical 
aerosol particles are tracked within the Large Eddy Simulation (LES) model. Since each 
Lagrangian particle on each time step can be assigned to the grid of 
the LES model model, based on 
location, forces due to condensation/evaporation of the water on these parcels
and drag forces
can be evaluated, and returned to the LES model. Although these two models 
\citeauthor{andrejczuk2008}~(\citeyear{andrejczuk2008},~\citeyear{andrejczuk2010})  
and \citeauthor{shima}~(\citeyear{shima})
differ 
in details, they both represent the mixed Lagrangian-Eulerian approach, with 
the microphysical process represented in Lagrangian framework.

Since the Lagrangian representation of microphysics in LES models is new, 
validation is required to ensure that the model is capable of producing 
results qualitatively and/or quantitatively comparable with observations.  
Such an evaluation is the main focus of this paper, using the VOCALS observations
to initialize and evaluate numerical model. 
This paper also shows the capability of the model to provide information about 
the relationship between aerosol and droplet properties inside a cloud that may help 
interpret aircraft observations or fill gaps in interpretation of these observations. 

The next section  
discusses the numerical model, initial conditions and the model set-up. Results are presented in section 3,
followed by conclusions in section 4. The appendix discusses the model sensitivity to the number of parcels
used in the simulations and the parameters the collision-coalescence algorithm depends on.
 
\newpage
\noindent {\bf 2. Numerical model} \\
The numerical model representing microphysics in Lagrangian framework
\citeauthor{andrejczuk2008}~(\citeyear{andrejczuk2008}, with the collision/coalescence process
described in \citeauthor{andrejczuk2010}~(\citeyear{andrejczuk2010}) is used to simulate a stratocumulus cloud. 
In this model microphysics formulated in a Lagrangian framework replaces the traditional Eulerian formulation and 
is coupled with the Eulerian 
dynamics and thermodynamics (\citeauthor{reisner2005},~\citeyear{reisner2005}).
The Lagrangian microphysics tracks millions 
of parcels, each representing aerosol/droplets with the same chemical and physical 
(location in space, velocity) properties with the information about the dynamical and thermodynamical 
conditions provided by the Eulerian model. For each parcel, a full condensation model is solved and
forces resulting from the phase change, together with the drag force are return to the Eulerian 
part of the model. Representation of the coalescence process in this framework 
calculates the collisions between each pair of a 
Lagrangian parcels within the collisional grid and creates new parcels. 
To limit the number of newly created parcels, a microphysical grid is used. Instead of creating a 
new parcel for each coalescence event, a new parcels are created for the microphysical grids 
for which number of assigned physical droplets is biggest than specified threshold level 
- $T_L$ (4000 in the simulations reported in this paper, which corresponds to
resolving 1 cloud droplet per m$^{3}$). This constraint is not sufficient, however, because
new parcels are created at each time-step and with time, the number of parcels grows. 
An additional constraint, is to limit the number of parcels within each microphysical grid to 4.
When there are more than 4 parcels in a microphysical grid, parcels are merged, with
the merging procedure designed to conserve number of droplets and the masses of
water and aerosol.
   
An important modification was introduced to the collision/coalescence to allow bigger 
time-steps between calls to collision procedure. When the 
time between collision calculations is too large, 
it can lead to a very small or even a negative
number of cloud droplets existing parcel represents after the collisions with all 
other parcels within the collision grid. 
To avoid this problem, a threshold level $T_L$ 
defining the minimum number of physical droplets within the Lagrangian parcel is also used to ensure that 
after a collision the number of physical droplets in the existing Lagrangian parcel is not less
than this threshold level. Additional setp is added in the collision algorithm to verify how many droplets
existing Lagrangian parcel would represent after the collision.
If this number is less than the threshold level, all collisions 
with this parcel are recalculated using the modified number of droplets this parcel represents - $N_i^{'t}$:
\begin{eqnarray}
N_i^{'t} = \frac{V_c}{\Delta t} \frac{T_L}{\sum_{j=1,j \ne i}^{P} K_{i,j}N_j^{t}},
\end{eqnarray}
where $V_c$ is the volume of the collision grid, $\Delta t$ is time between collisions, $T_L$ is a threshold level, 
$K_{i,j}$ - is the gravitational collision kernel, and $j$ indexes the Lagangian parcels within the collision grid.
After inserting this number into the equation describing cloud droplet number change after collisioni we have: 
\begin{eqnarray}
N_i^{t+1} = \frac{\Delta t}{V_c}N_i^{'t} \sum_{j=1 // j\ne i}^{P} K_{i,j}N_j^{t}
\end{eqnarray}
that is, Lagrangian parcel $i$ represents exactly $T_L$ cloud droplets at time $t+1$. 

In all the simulations reported in this paper, the described modifications were present. 
On some occasions this part of the algorithm was activated even in simulations calling 
the collision procedure every time-step. It produced 
negligible effect on solution. Because the collision algorithm has big impact on the 
execution time, calculation of the collision procedure with the time-step bigger than $dt$ has advantages. 
However, as discussed in the appendix, modification to collision algorithms may also have significant 
impact on execution time when the time interval between calls to collision algorithm is too big. 

\noindent {\bf 3. Model set-up and initial conditions} \\
All the simulations reported here are in 2D. The domain covers 3200m in the horizontal
and 2000m in vertical, and is resolved with the 80$\times$200 points.
All the simulations were run for 7 hours with time-step 0.2s. For the first 2 hours the models were run without
the collision process.
The Eulerian microphysical grid for the collision mapping covers 1 - 500 ${\mu}m$ in the cloud droplet 
radius space and 0.005 - 5 $\mu m$ in
the aerosol space and is resolved with 30$\times$30 grids. 

To derive the initial conditions for the model, for each of the 3 discussed cases, a cloud penetration,
long enough to supply information about the dynamical, 
thermodynamical, microphysical profiles and aerosol distribution
through the boundary 
layer was chosen from the VOCALS observations (\citeauthor{Wood2010},~\citeyear{Wood2010}). 
Three cases, refered as HIGH (profile from 13 Nov. 2008),
MED (profile from 31 Oct. 2008) and LOW (profile from 13 Nov. 2008) were chosen  based on the
cloud droplet concentration measured inside the cloud.  
For all 3 cases profiles of potential temperature ($\theta$) and water vapour 
mixing ratio ($q_v$) were specified as:

\begin{equation}
\theta(z)=
\begin{cases}
   {\theta}_B,                  & z \le {z_B};\\
   {\theta}_C+{\alpha}z,           & z_B < z \le z_T;\\
   {\theta}_T+(z-z_T)^{2.8},   & z > z_T;
\end{cases}
\end{equation}
\begin{equation}
 q_v(z)=
\begin{cases}
  q_{vB}~\text{(or saturation)}  &\text{if $z \le z_T$;}\\
  q_{vT}                  &\text{if $z > z_T$;}\\ 
\end{cases}
\end{equation}
with constants for each simulation defined in table \ref{table.profiles}.

Radiative forcing is based on the \citeauthor{bstevens1}~(\citeyear{bstevens1})
parametrization. Only radiative
cooling at the cloud top is taken into account and the longwave radiative flux
is expressed as:
\begin{eqnarray}
F_{rad}=F_0 exp^{-\kappa \int_z^{\infty} \rho q_c dz}
\end{eqnarray}
with $\kappa=85 m^2/kg$ and $F_0=35 Wm^{-2}$.

\begin{figure}[htb]
\begin{center}
\epsfig{figure=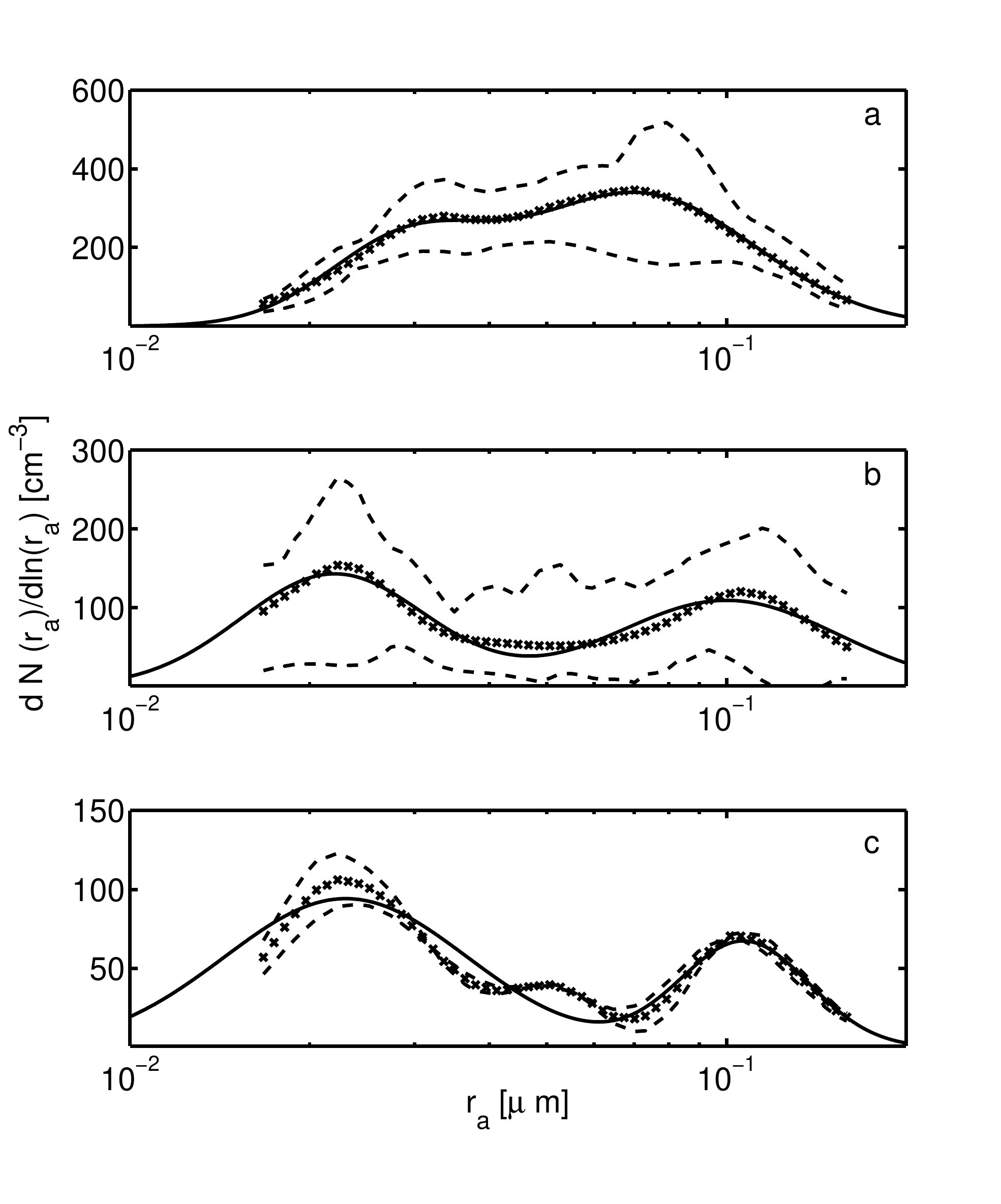,height=5in,angle=0}
\end{center}
\caption{
Averaged measured (x), fitted(solid line) and maximum/minimum measured (dashed line) of aerosol
distribution for HIGH (a), MED (b) and LOW (c) cases.
}
\label{tmpfigure1}
\end{figure}

For all cases the initial cloud water mixing ratio was specified, varying linearly with height
inside the cloud,
with the maximum value at the top of the boundary layer of $0.32 [g/kg]$. 
The horizontal velocity (u component) was specified as height independent and was taken as 
the maximum of $u$ and $v$ from the observations:  $1 [m/s]$ for the HIGH run, $7 [m/s]$ for the MED run 
and $4 [m/s]$ for the LOW run. For all 3 cases the aerosol spectrum from SMPS 
(Scanning Mobility Particle Sizer) 
were analysed 
and a two modal log-normal distribution was fitted to the averaged aerosol spectrum. 
For the HIGH and MED cases all observations from below the cloud 
within time specified in table \ref{table.profiles} 
were used to calculate 
the parameters of the distributions. 
Averaged observations (x in figure \ref{tmpfigure1}), maximum and minimum values measured 
for each SMPS bin (dashed line) 
and a fitted 
log-normal distribution (solid line) for these 2 cases are shown in 
figure \ref{tmpfigure1}a and b. For the LOW case averaging all the observations 
below the cloud led to a distribution which for given forcing produced a much higher than 
observed cloud droplet concentration. Out of the 7 aerosol spectrum measurements only two,
 just below the cloud were taken into account when fitting log-normal distribution, 
with the resulting distribution shown in figure \ref{tmpfigure1}c. 
Parameters of a log-normal distribution for each of the cases 
are shown in table \ref{table.lognormal}. Additionally log-normal distributions were fitted
to the $mean$+/-${\sigma}$ and coefficients are also included in the table. For LOW case
table \ref{table.lognormal} shows also coefficients for the whole below the cloud measurements
including $mean$+/-${\sigma}$.   

Cases HIGH and MED are similar, with the observations taken in the region of 
unbroken cloud. Despite the fact that the flights were on different days, the height of 
the boundary level was about \~ 1400m for both cases, 
cloud is thicker for the HIGH case and is about 580 meters deep, 
compared to 400 meters for the MED. The biggest difference is in the cloud droplet concentration, with 
250 cloud droplets $cm^{-3}$ for the HIGH run and 120 cloud droplets $cm^{-3}$ for the MED run. The 
LOW case is different. Although measurements were taken on the same day as for HIGH run, 
the boundary layer height is less than 1300 meters, cloud thickness is 
360 meters and cloud droplet concentration is about 65 $cm^{-3}$ 
The satellite image (e.g. \citeauthor{andrejczuk2011},~\citeyear{andrejczuk2011} figure 10) 
shows, that the region where measurements were taken is at the 
edge of the unbroken cloud with the circulation changing from closed to open cells to the west of this region.

In all the simulations 100 parcels per grid cell were used to represent the initial aerosol distribution,
resulting in 1.6 million parcels initially. As the model creates new parcels in the coalescence process
the simulation HIGH, MED and LOW ended with 2.5, 2.3 and 2.2 million parcels respectively.
 
\begin{figure}[htb]
\begin{center}
\epsfig{figure=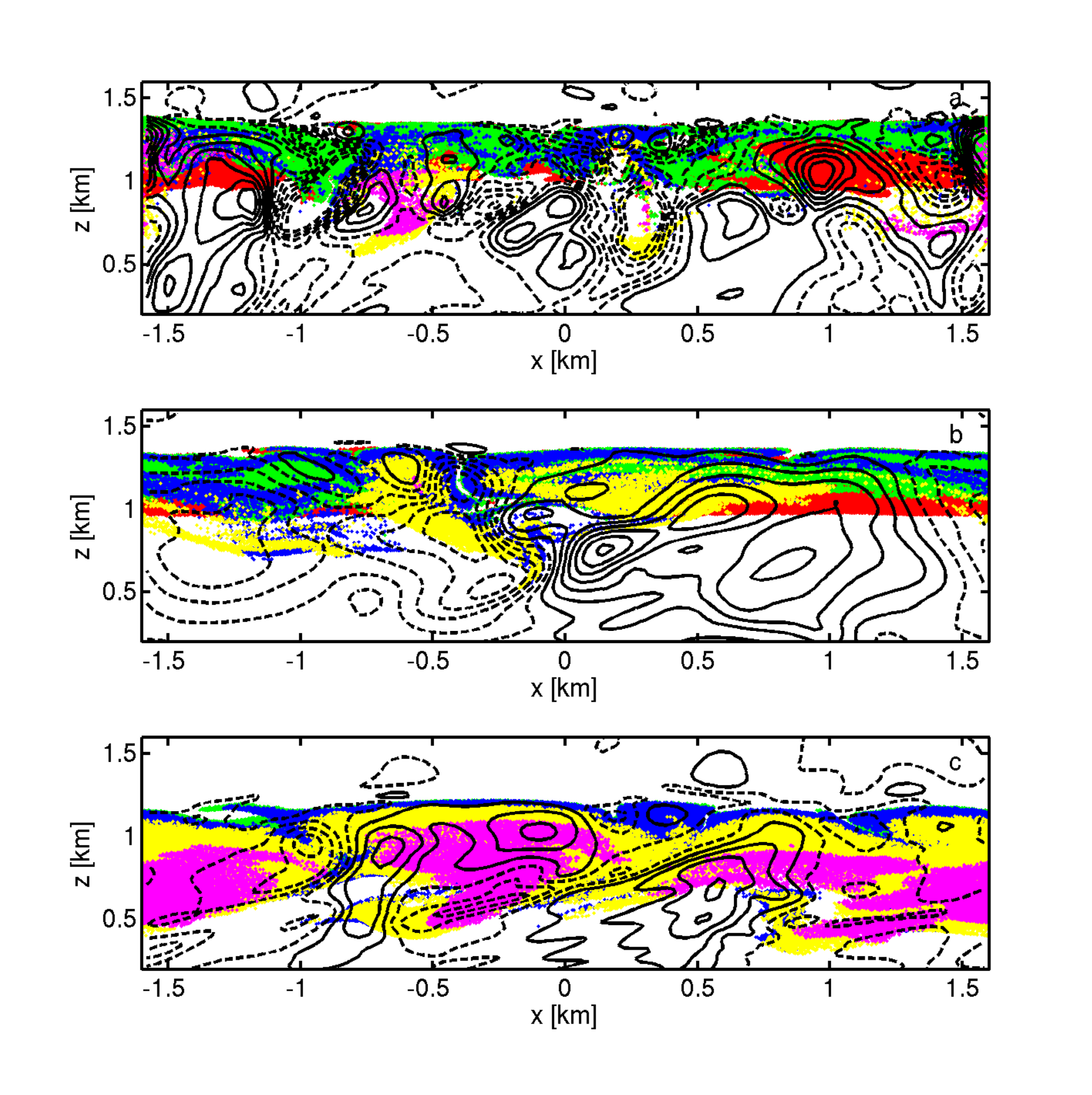,height=5in,angle=0}
\end{center}
\caption{  
Locations and sizes of Lagrangian parcels after 7th hour of simulations; 
run HIGH - panel a, run MED - panel b, run LOW - panel c. Red colour parcel sizes 1-10 ${\mu}$m,
green - 10-20 ${\mu}$m, blue 20-50 ${\mu}$m, yellow 50-100 ${\mu}$m, and magenta - bigger than 100 ${\mu}$m.
}
\label{part_loc}
\end{figure}

\noindent {\bf 4. Results}\\
\noindent {\bf 4.1 General properties}\\
Lagrangian representation of the microphysics provides information about the Lagrangian parcels locations and the sizes
for each time step. This is shown in figure \ref{part_loc}, after 7 hours, 
with each Lagrangian parcel with a radius
bigger than 1 ${\mu}m$ assigned to one of four classes with sizes [1-10] (red), [10-20] (green), [20-50] (blue), 
[50-100] (yellow) and bigger 
than 100 ${\mu}m$ (magenta). Additionally contours of velocity are shown with 
solid line for positive and dashed line for negative values every 0.1 m/s, with contour 0 omitted. 
Figure \ref{part_loc} shows the complex 
structure of the flow and the parcels distribution in space. Typically the biggest drops (magenta) are located near the cloud 
base and centre and smallest droplets are distributed throughout the whole cloud. In some areas, in the up-draft, for the MED case,
there is a clear transition from 
small to big droplet sizes (red-$>$green-$>$blue), but very often entrainment and falling droplets creates mixtures 
of the green and blue; the range of sizes droplets can grow by condensation. 

\begin{figure}[htb]
\begin{center}
\epsfig{figure=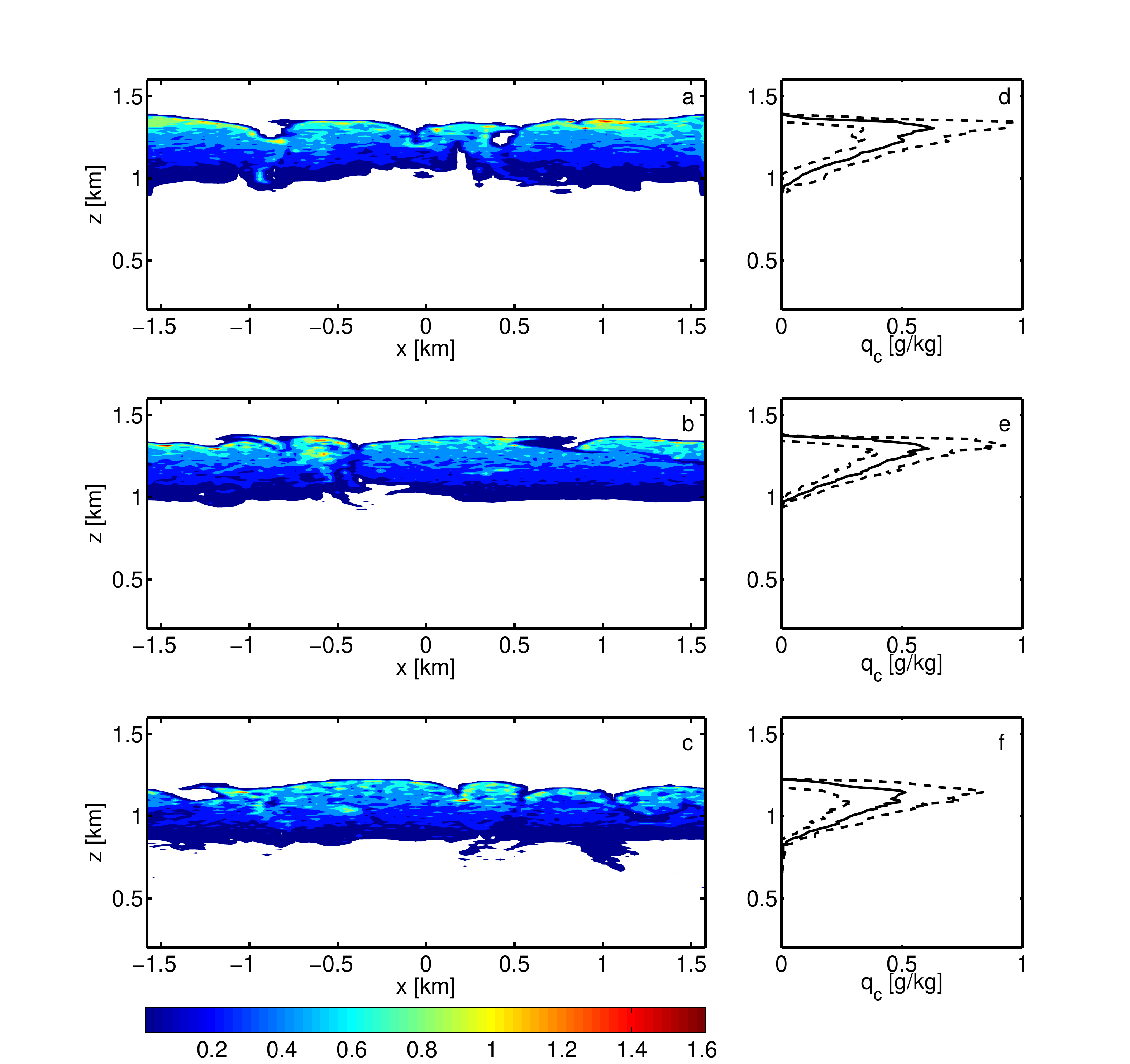,height=5in,angle=0}
\end{center}
\caption{  
Panels a-c cloud water mixing ratio diagnosed from Lagrangian parcels location 
and size after 7 hours of simulations. Panels d-f - cloud water mixing ratio profile (solid line),
with standard deviation (dashed lines). 
}
\label{qc_contour}
\end{figure}

Since each parcel can represent a different number of cloud droplets/aerosol particles, with
the biggest, created in collision process shaving the smallest number, the fact that parcels are present
in the model grid does not necessary mean that there is a significant amount of water in this grid. Combined information
about size and number determines total mass of water in each grid. The cloud water mixing ratio diagnosed from the parcels
size and location is presented in figure \ref{qc_contour}a-\ref{qc_contour}c. Only values bigger than 0.05 g/kg are 
plotted with an interval of
0.1 g/kg. The cloud top reflects features shown in figure \ref{part_loc}, since there are almost no parcels with radii 
bigger than 50 ${\mu}m$, created in the coalescence process and as a result representing much less cloud droplets than
were initially assigned to each parcel. Near the cloud base, where such parcels are present, the contour of 0.05 g/kg is
around 1000m for HIGH and MED and 800m for LOW. At the same time there are parcels present below these levels. These
however, despite their big sizes, represent small number of cloud droplets and contribute little to $q_c$. 
Panels \ref{qc_contour}d-\ref{qc_contour}f show
mean profiles and standard deviations of $q_c$.
The standard deviations of the cloud water mixing ratio are big, not only 
because of the variability
due to dynamics, but also due to sampling error.
The latter is a result of representing aerosol distribution with a limited
number of parcels - 100 per grid initially. The transfer of one parcel from one grid to another represents a
transfer of 1\% of total the number of aerosol - leading to $q_c$ variability since the same number of parcels in each grid
can not be guaranteed.

\begin{figure}[htb]
\begin{center}
\epsfig{figure=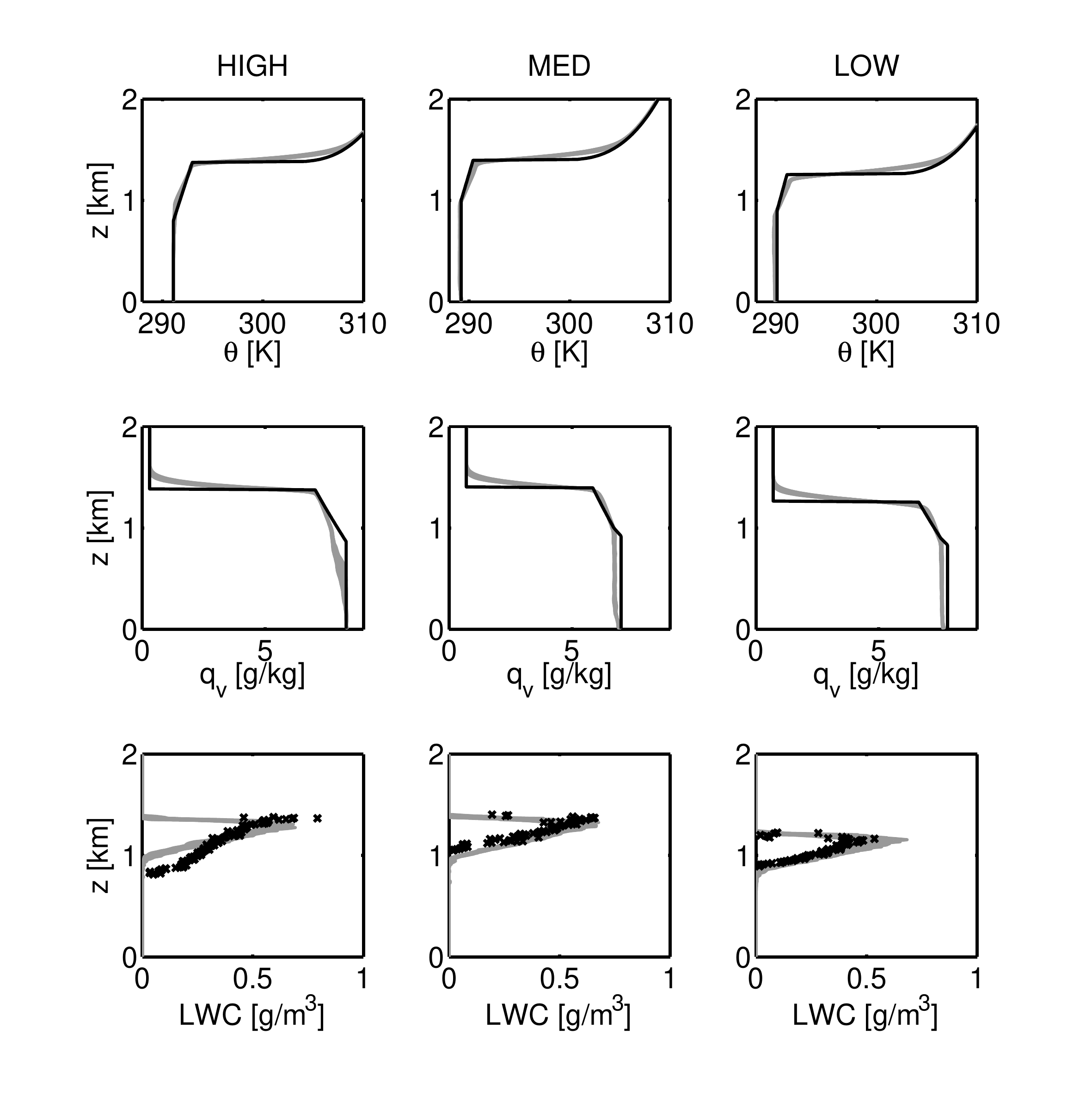,height=6in,angle=0}
\end{center}
\caption{  
Initial profiles of $\theta$ and $q_v$ - black line and model solution for HIGH run (a and b), MED run (d and e)
and LOW run (g and h) - grey lines. Panels c, f, i - LWC profiles derived from model (grey lines; last 3 hours) 
with observations (CDP probe) - x.
}
\label{thermodynamicalprofiles}
\end{figure}

The initial profiles and model solutions for potential temperature and water vapour mixing ratio are shown on
figure \ref{thermodynamicalprofiles} together with observed and modelled LWC (Liquid Water Content). 
The black lines show initial profiles 
derived from observations, and the grey lines show model output for the last 3 hours with the output
saved every 6 minutes. For the LWC, observations rather than initial profiles are shown with each point 
representing one sample from the CDP (Cloud Droplet Probe) probe.  
Both $\theta$ and $q_v$ profiles show departure from initial 
values. The numerical model is unable to maintain sharp gradients near the top of the boundary layer,
and below the cloud layer is cooling and drying with time.

Profiles of LWC - figure \ref{thermodynamicalprofiles} show little change with time 
for all simulations, with 
the profiles for last 3 hours shown with the grey lines and observed profiles derived from the 
CDP measurement with $x$. 
Out of 3 cases considered in this paper the best agreement 
in LWC profiles occur for LOW and MED simulations, with slightly deeper cloud produced by model for both runs. 
The HIGH simulation shows much bigger differences between model and observations especially near the cloud 
base. However, for MED and LOW cases, the LWC change with height 
is close to linear, whereas for  HIGH case 
there is a change of slope near the cloud base, and may result from larger scale perturbation 
in the observations, not
resolved by the model due to limited domain. Clearly, the  model, when starting from an idealized $q_c$ 
profile, is not able to reproduce this feature in the solution.  

\begin{figure}[htb]
\begin{center}
\epsfig{figure=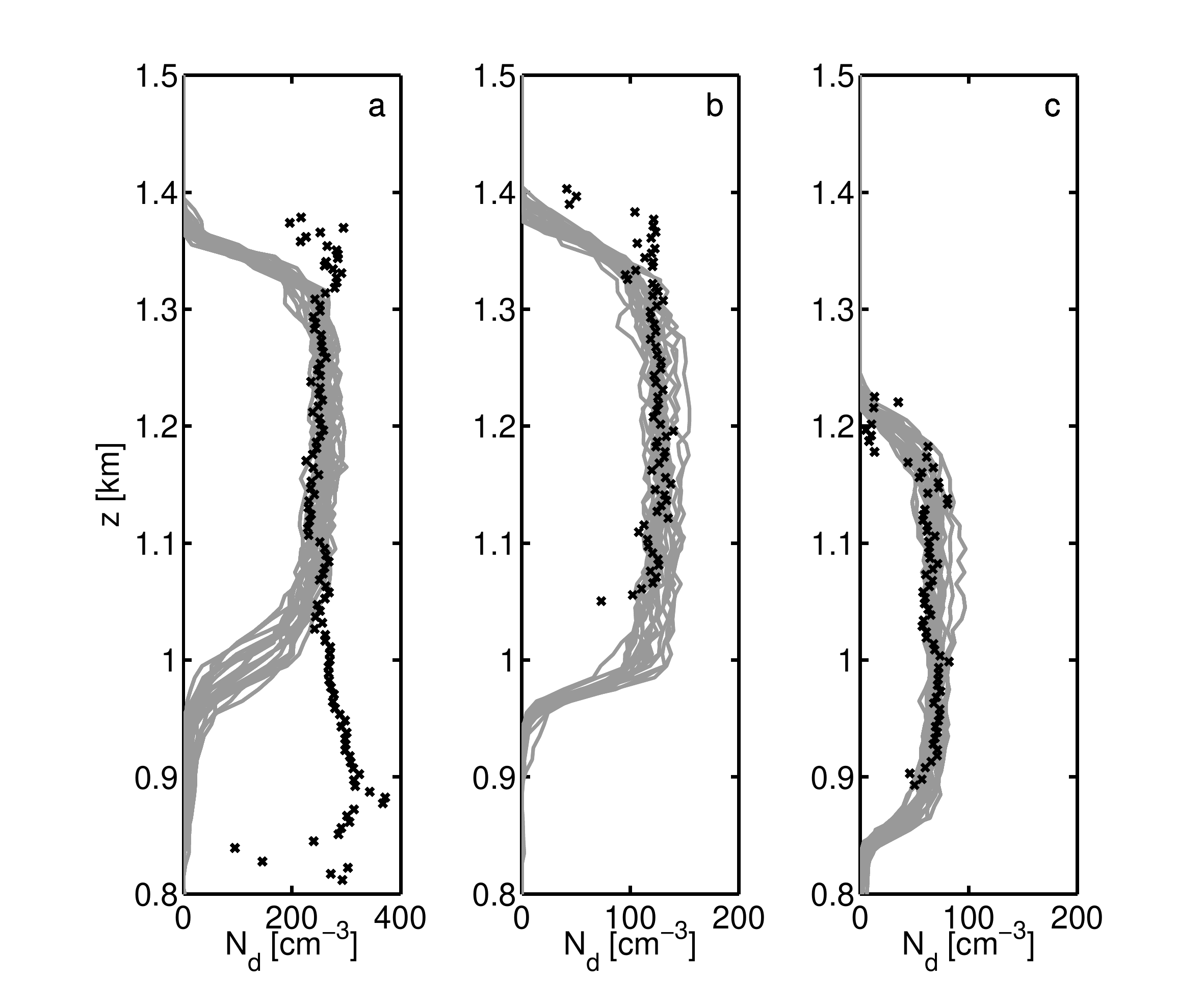,height=6in,angle=0}
\end{center}
\caption{  
Profiles of cloud droplet number as measured  by CDP probe (x) and modelled (grey lines; last 3 hours) 
for runs HIGH (a), MED (b), LOW (c).
}
\label{CDNprofile}
\end{figure}

\noindent {\bf 4.2 Microphysics}\\
\noindent {\bf 4.2.1 Cloud droplets}\\
The cloud droplet number profiles measured by the CDP probe for each cloud section ($x$) and 
the corresponding model profiles are shown on 
figure \ref{CDNprofile}. Although later the observations from CDP and 2DS (Two-Dimensional Stereo probe) 
are combined to produce cloud droplet spectrum, the
droplet concentration from the 2DS probe for radii bigger than 25 ${\mu}m$ was less than 2 cm$^{-3}$ and these not included 
in figure \ref{CDNprofile}. 
For the numerical model the cloud droplet concentration was calculated based on the
particle radius, with all particles with sizes bigger than 1 ${\mu}m$ included. 
This method was 
used, instead of taking only cloud droplets having radius bigger than activation radius, because the CDP probe
can not make distinction between cloud droplet and non-activated aerosol. 
In the numerical model, since full information about each parcel and thermodynamics is available
such a distinction can be included  
and this is discussed later.
For all 3 cases, the cloud droplet concentration changes little with height for both, simulations and observations.
This suggests, that there is either a balance in droplet activation in down-drafts and deactivation in up-drafts,
or droplets activate/deactivate near the cloud base only.
The agreement in the cloud droplet concentration between the model solution and observations for all 3 runs 
is very good. Model profiles show the variability with time. 
This variability is, however, relatively small 
and are expected since the numerical solution is not in a steady state, but varies with time because the cloud may be at different 
stage of the development for each saved time. 
As for the LWC, cloud droplet number profiles indicate that model is producing 
a deeper cloud, for MED and LOW cases. For the MED case, the model also under-predicts cloud top height.  
For the HIGH run agreement between observations and model solution is not as good as for two other cases. 
As discussed earlier, the model does not capture the structure 
observed below 950 m.,  
and under-predicts water content and cloud droplet number there, when starting from 
idealized profiles. 

\begin{figure}[htb]
\begin{center}
\epsfig{figure=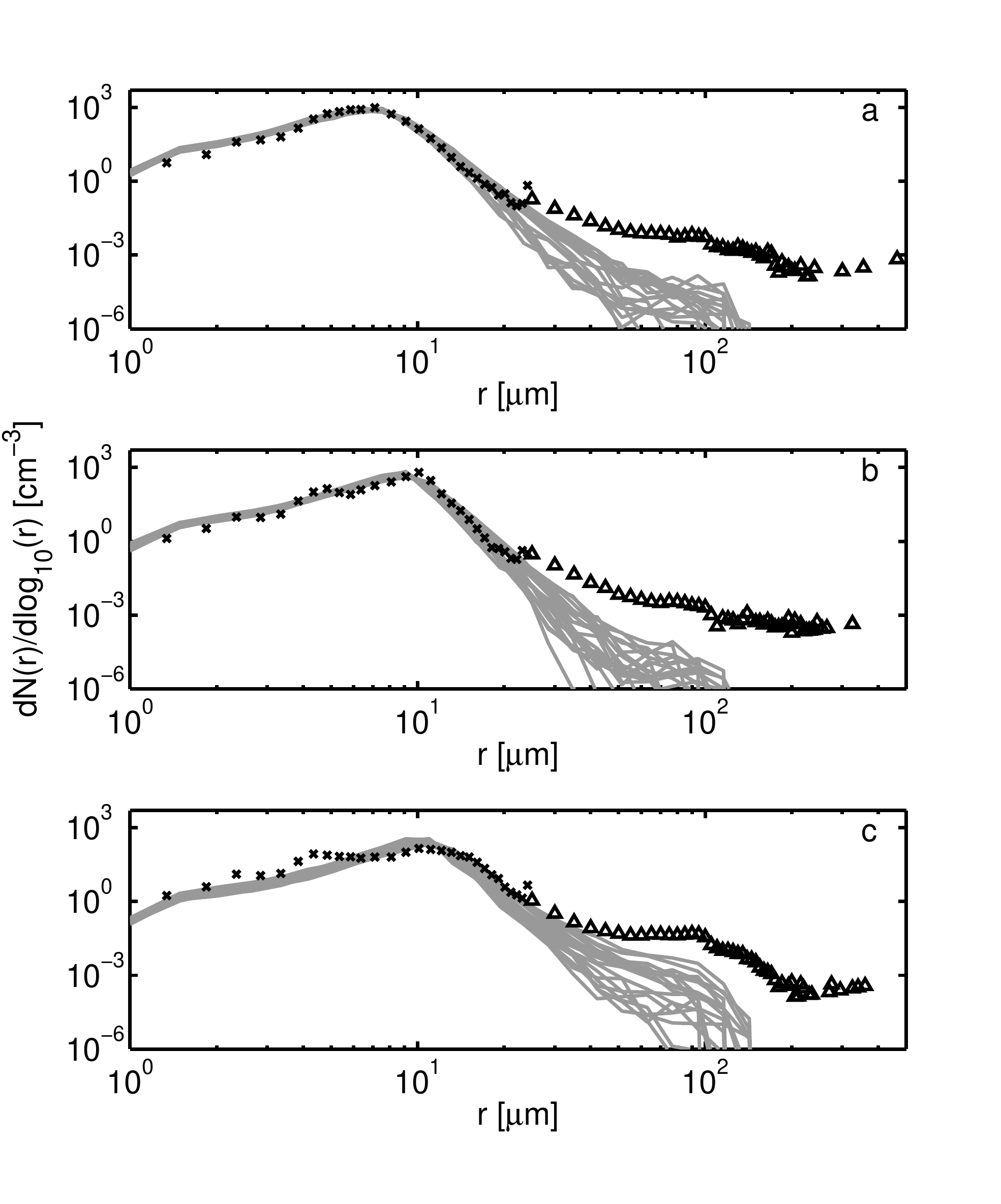,height=7in,angle=0}
\end{center}
\caption{  
Cloud droplet spectrum averaged over the whole domain. 
CDP probe (x), 2D-S probe triangles and from model grey lines (last 3 hours).
}
\label{qrspectrum}
\end{figure}

For the BAe-146 flights, the CDP and  2D-S were the only probes for which the droplet spectra were available. 
The CDP probe samples droplet with radius between 1 and 25 $\mu m$ and thus provides mainly information about 
cloud droplets. The 2D-S provides information about droplet sizes larger than 2.5 ${\mu}m$, but only
sizes bigger than 25 $\mu m$ were used for the analysis.  
Figure \ref{qrspectrum} displays the CDP measurements (with x), 2D-S measurements (with triangles) and model solution 
(each line represents one time with spectrum averaged over the whole domain for the last 3 hours) 
for passes
through the cloud for the 3 cases considered in this paper. Only sizes smaller than 500 ${\mu}m$, the biggest bin size
in the model, are shown. 
Spectra from the observations and from the model show good agreement for cloud droplets with radius r $<$ 25 ${\mu}m$ 
for HIGH and MED cases. For the LOW case agreement is not as good, but still reasonable. 
For the droplets larger than 25 ${\mu}m$ (measured with 2D-S) the model under-predicts the droplet concentration for all 
3 cases. Now, however, for the LOW case, the model solution is closest to the observations. 
Under-prediction of the biggest droplets concentration in the model
indicates that
coalescence process in the model is not efficient enough in production of big droplets. 
However
it should be noted that there is a difference in the cloud droplet concentration 
for CDP and 2D-S for radius 25 ${\mu}m$, with the 2D-S measuring higher than CDP concentrations (especially for
HIGH and MED). It may indicate that the 2D-S over-predicts the concentration of the big droplets or that the 
sample was not big enough to produce the right statistics.  
There is no consistent trend in the concentration of the largest drops. Both the observations and model
show highest drop concentrations for the 
LOW case and lowest for the MED case. 

\begin{figure}[htb]
\begin{center}
\epsfig{figure=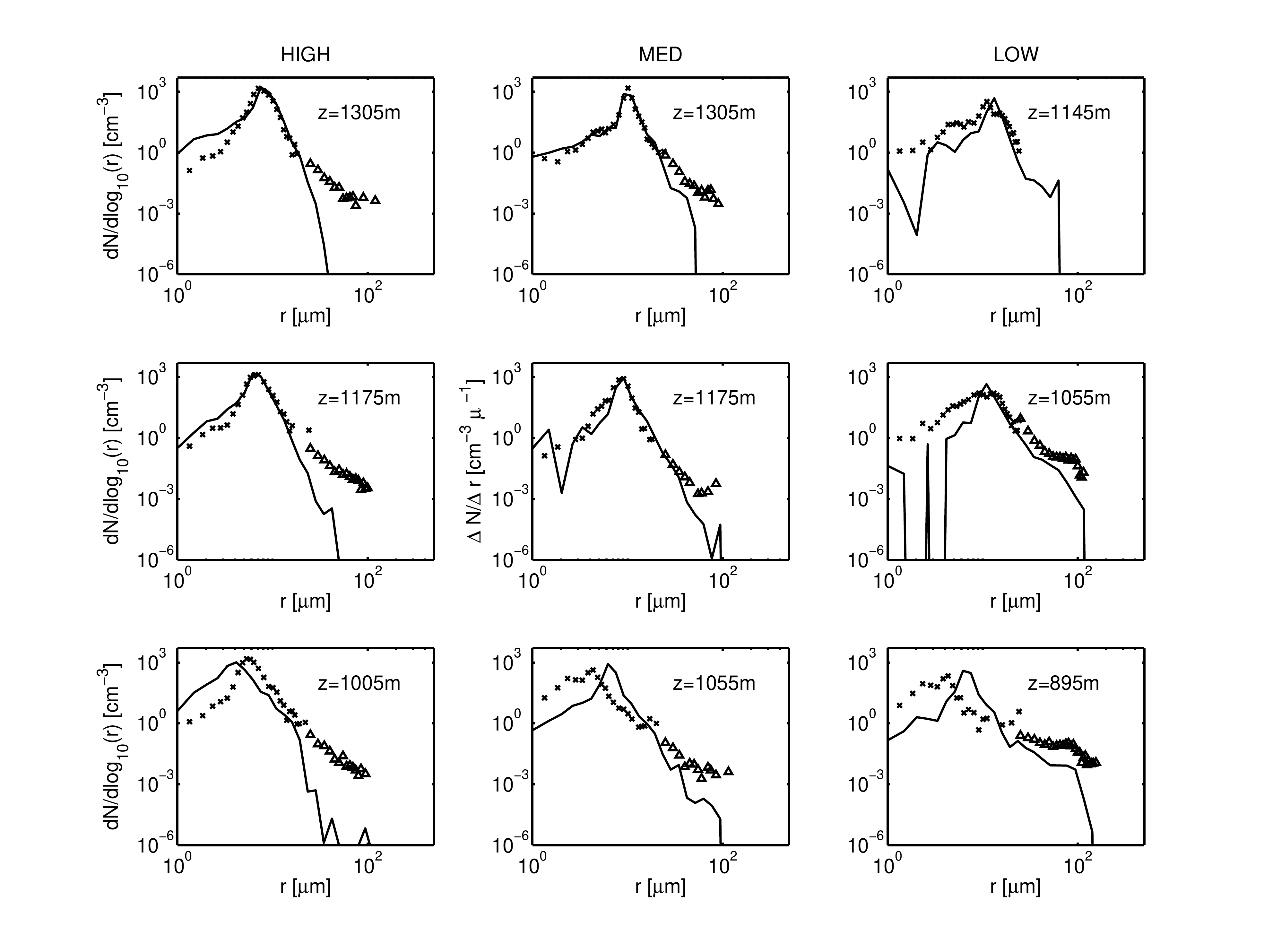,height=5in,angle=0}
\end{center}
\caption{  
Cloud droplet spectrum for different locations inside the cloud. Top row - near the cloud top,
middle row - near cloud center, low row - near the cloud base for HIGH (left column),
MED (middle column) and LOW (right column). Solid line - model, symbols - CDP (x) and 2D-S (triangles)
}
\label{spectrumdetails}
\end{figure}

Although averaged over the whole domain the spectra from the observations and the numerical model are similar, 
a detailed analysis shows that there are 
differences in droplet spectrum when comparing model - 7 hours into the simulation 
and the observed spectrum at a given level.
The droplet spectrum for 3 different heights in the cloud are shown in figure \ref{spectrumdetails}, with the top 
row showing spectrum near the the cloud top, the middle row showing spectrum 
in the middle of the cloud, and the bottom row showing spectrum near the cloud base for HIGH, MED and LOW runs. 
For the numerical model the spectrum on any given vertical level was horizontally averaged on this level for 
grids where $q_c > 0.001 g/kg$. 
From the observations the spectra at the height $z_i$ were obtained 
by averaging the measurements in the vertical between $z_i-.5\Delta z$ and $z_i+.5\Delta z$, 
with $\Delta z =10m$ - model grid size. Figure \ref{spectrumdetails} shows that the model has 
problem with capturing details of the droplet spectrum on a given level. The differences between the model
and the observations are present not only for the biggest droplets, created in the coalescence process, but also 
for the smallest. 
As the numerical model solution depends not only on microphysics, but also on
the dynamics of the flow, this suggest that the dynamics of the solution may not be captured correctly 
perheps due to the lack of large scale forcing and the simulation
being 2 dimensional. 
Note, however, that
the observations were taken from only one penetration through the cloud for each case. 
Since the aircraft instrument samples only a small 
volume of the atmosphere, this disagreement may also indicate insufficient resolution of the observations. 

\begin{figure}[htb]
\begin{center}
\epsfig{figure=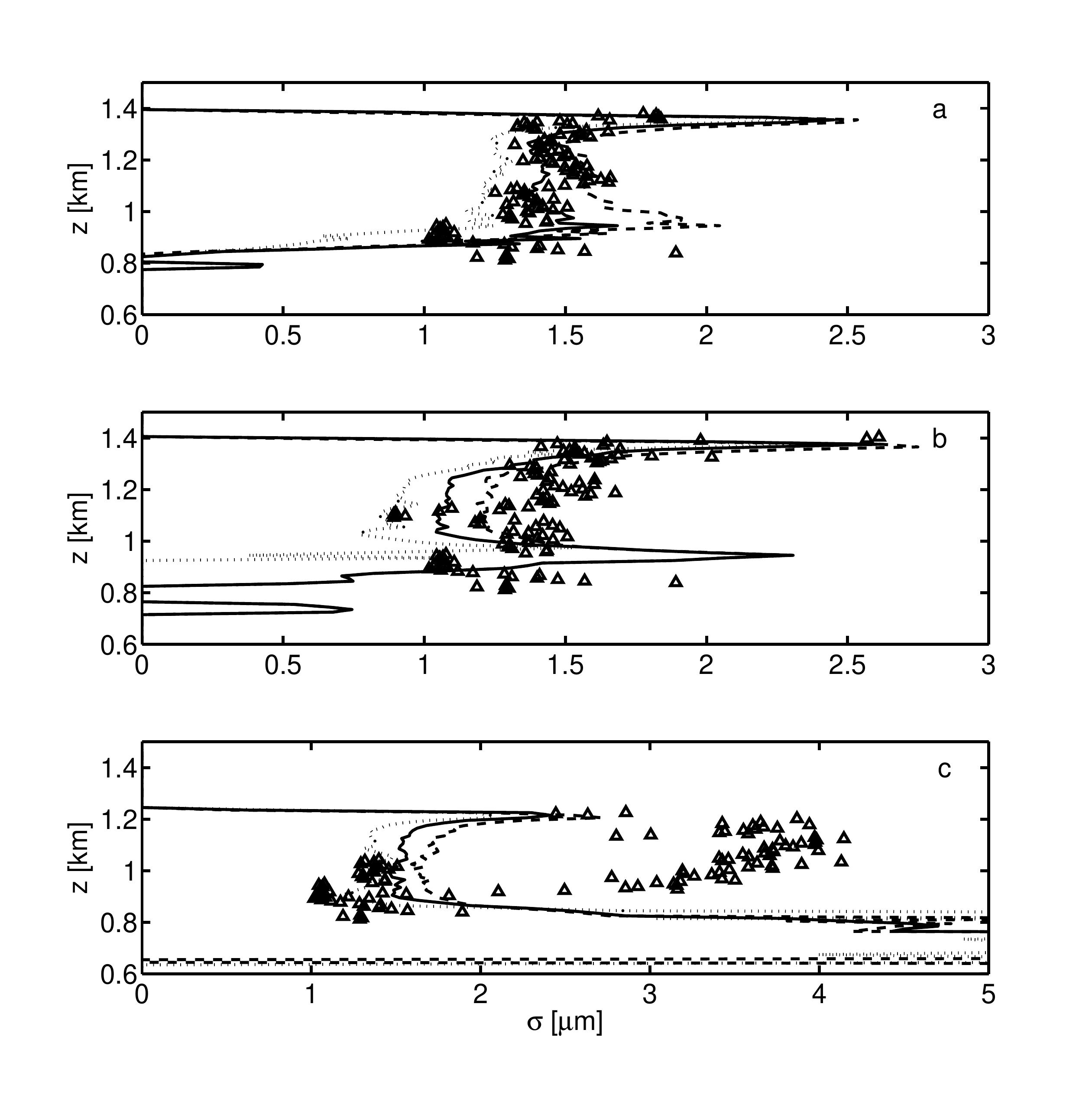,height=7in,angle=0}
\end{center}
\caption{  
Profiles of the standard deviation of the cloud droplet distribution, measured by CDP probe (x) and modelled - lines.
Solid line - whole cloud, dotted line up-drafts only, dashed line down-drafts only. 
Run HIGH - panel a,
run MED - panel b and run LOW - panel c
}
\label{sigma}
\end{figure}

The width of the droplet spectrum is an important factor affecting the coalescence process. Parcel models solved for 
the growth of an individual droplets tend to produce very narrow spectrum. 
Despite the fact 
that the condensation model used in the LCM is formulated in a Lagrangian framework, it does not produce 
narrow spectrum and the coalescence process is actively producing big cloud droplets. The latter has been shown
already in figure \ref{part_loc} where big  droplets are present even for the run with a high aerosol 
concentration. Figure \ref{sigma} shows profiles of the standard deviation ($\sigma$) of the cloud 
droplet distribution averaged over the last 3 hours of the simulation for the 3 cases considered in this paper, 
with the standard deviation showing values around 1.4 $\mu m$ for HIGH, 1.1 $\mu m$ for MED and 1.5 $\mu m$ for LOW run, 
and varying 
with the height in the cloud. Only standard deviation smaller than 3 ${\mu}m$ 
(5 ${\mu}m$ for LOW run)
are shown since near the cloud base fast evaporation for small droplets and slow for  big droplets leads to 
significant growth of the standard deviation.  Near the cloud top the increase of the standard deviation up to 3 $\mu m$ 
is observed as a result of entrainment, mixing and subsequent evaporation. Aside from $\sigma$ for the 
whole cloud, values for the up-draft and down-draft are also displayed. The general picture 
is similar for all 
runs, with $\sigma$ in down-drafts bigger (as a result of droplet spectrum broadening during evaporation) 
than in up-drafts (droplet spectrum narrowing due to condensation). There is a tendency for an increase in $\sigma$ 
in down-drafts near the cloud base as can be expected in the region of evaporation. 
No significant narrowing of the cloud 
droplet spectrum is observed in up-drafts (aside at the cloud base), 
contrary to parcel model predictions, where such a tendency is 
present [\citeauthor{bartlett},~\citeyear{bartlett}, \citeauthor{warner},~\citeyear{warner}]. 
There are, however, significant differences between the microphysics in a parcel model and in 
the LCM model. In the LCM each parcel travels along different trajectory 
and encounters a different supersaturation history along the trajectory; dynamic and thermodynamic parameters 
are interpolated to the parcel location to determine the supersaturation for each parcel and as a result 
the effect of neighbouring grids 
and variability of the supersaturation inside the grid 
is taken into account; parcels do not move along air trajectories, with the velocity equations solved for each 
parcel; and the coalescence process is very efficient in 
broadening the spectrum. Some of these processes and their effect on spectrum broadening were theoretically
investigated by \citeauthor{cooper}~(\citeyear{cooper}) and \citeauthor{srivastava}~(\citeyear{srivastava})
in the context of the turbulence and microscopic fluctuations in the supersaturation. Note, however, that in the model
variability does not derive from these processes, 
but are purely a numerical effect because the dynamic/thermodynamic parameters are
interpolated to the parcel location. 
The numerical model shows no consistent 
trend in  $\sigma$ behaviour for the whole cloud as a function of aerosol/cloud droplet concentration.  
It is biggest for the LOW and HIGH runs and smallest for MED run.

Additionally, figure \ref{sigma} indicates the standard deviation of cloud droplet 
distribution derived from the CDP probe by means of triangles. Good agreement between the observations and the model is observed
for HIGH and MED runs. For the LOW run the model under-predicts the standard deviation of the droplet spectrum 
inside the cloud by a factor of more than 2, with the 1.5 ${\mu}m$ modelled and 3.3 ${\mu}m$ observed. 
Few data points with the standard deviation smaller than the modelled values are 
also present in the observations for this case, 
indicating significant variability of the standard deviation in the observations.
This difference between the model and the observations for the LOW case may be a result of the two dimensional domain. 
\citeauthor{andrejczuk2010}~(\citeyear{andrejczuk2010}) 
shows a $\sigma$ of almost 4 ${\mu}m$ in 3D simulations using aerosol distribution 
observed during DYCOMS for very low aerosol concentrations.

\begin{figure}[htb]
\begin{center}
\epsfig{figure=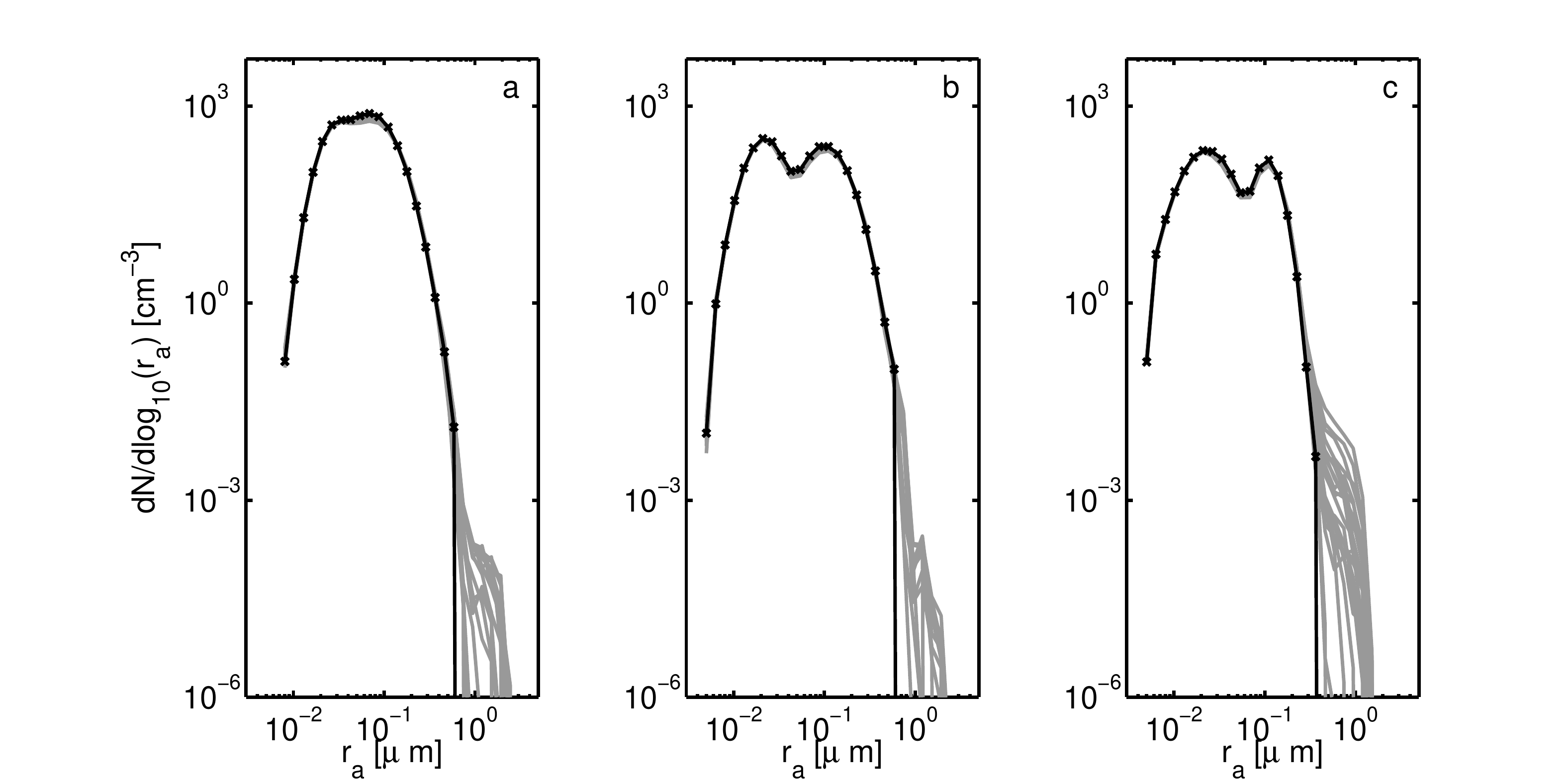,height=3in,angle=0}
\end{center}
\caption{
Evolution of the aerosol distribution for the last 3 hours (grey lines) and initial aerosol distribution black line with 'x'
for run HIGH - panel a,
MED - panel b and LOW - panel c.
}
\label{aerosol}
\end{figure}

\noindent {\bf 4.2.2 Aerosol} \\
Understanding the aerosol-cloud droplet relation is important not only for climate models, but also for
weather prediction models. Due to insufficient resolution, these models can not capture the variability 
within the cloud and as a result require parametrization 
of the relation between the aerosol concentration and the cloud droplet concentration.
Results in this paper show that model with Lagrangian
representation of microphysics gives a good agreement with observations not only for the cloud droplet 
number but also for the cloud droplet spectrum. 
In the coalescence process implementation, where 2 dimensional grid is 
used to map result of collision, both the aerosol radius and the cloud droplet radius is changing as a result of parcel
collisions. With time, not only is the radius of the cloud droplets growing, 
but also the size of the aerosol inside this droplet. 
Figure 
\ref{aerosol} shows the aerosol spectrum inside the cloud, with 
initial aerosol spectrum (solid line), and spectra for the last 3 hours every 6 minutes 
(grey lines). For all cases, the model is producing sizes larger than the initial distribution as a result of droplets 
coalescence, with the biggest aerosol radius about 2.3 ${\mu}m$ for the HIGH and MED runs and 1.5 ${\mu}m$ for the LOW run. 
For the LOW run, since the aerosol concentration is smallest and the droplet sizes biggest, one can expect 
to see larger maximum aerosol sizes than for the HIGH and MED runs. However one should appreciate that the biggest initial 
aerosol size was about 0.5 ${\mu}m$ for the HIGH and MED and about 0.2 ${\mu}m$ for the LOW case. 
It follows that the aerosol size grew almost 5 times in the HIGH and MED cases and almost 8 times in the LOW case.  
Production of the big aerosol particles in the collision process is rather slow and after 5 hours for the LOW run, only
about 100 aerosol particles per m$^3$ with radius 1 ${\mu}m$ is present inside the cloud.

\begin{figure}[htb]
\begin{center}
\epsfig{figure=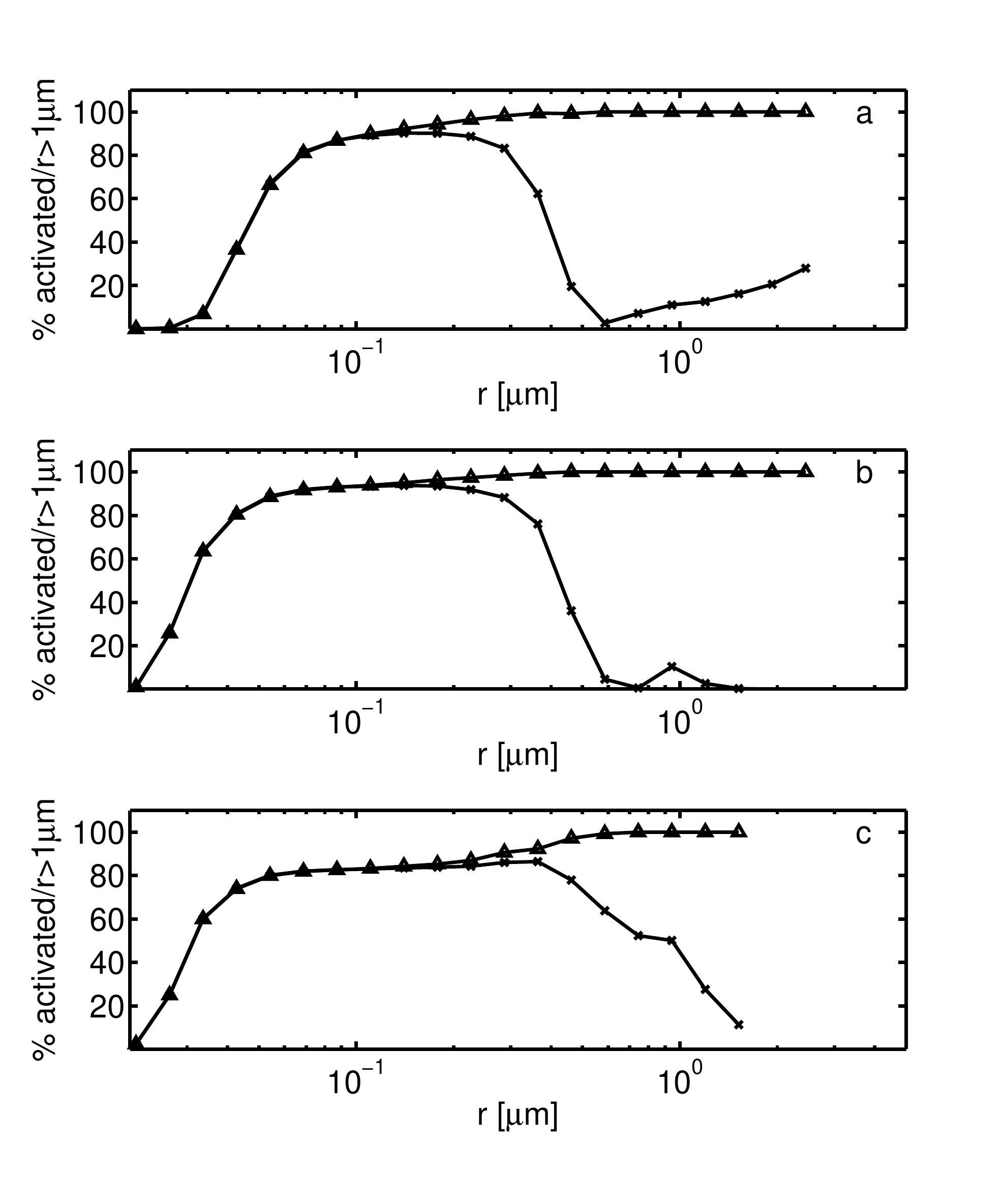,height=7in,angle=0}
\end{center}
\caption{  
Percentage of droplets with radius bigger than the activation radius (stars), and with droplet size bigger than 1 
${\mu}m$ - both activated and not activated (triangles) as a function of aerosol radius for run HIGH - panel a,
MED - panel b and LOW - panel c. Averages over the last 1 hour were taken to calculate the statistics.
}
\label{actaerosol}
\end{figure}

Since the model gives detailed information about the aerosol and cloud droplets, it can be used to produce 
a diagnostic relation about aerosol activation inside the cloud.
Two lines are shown on each panel of the figure \ref{actaerosol} with the averages over the 
last 1 hour taken to calculate the statistics. 
The solid line with $x$ shows the percentage 
of activated aerosol - with a radius bigger than activation radius for each aerosol bin. 
For all cases the aerosol must be at least $0.02 {\mu}m$ to be activated. The percentage 
of activated aerosol increases with the aerosol radius up to 90 \% for the HIGH and MED run or 85 \% 
for the LOW run and then decreases significantly. 
On the same plot a line with triangles shows the percentage of aerosol particles inside
the droplets a with radius bigger than $1 {\mu}m$  
(both activated aerosol and aerosol with water on it but not exceeding the activation radius).  
This line shows a constant 
increase to 100 \% reached for aerosol sizes of about $0.3 {\mu}m$ for the HIGH and 
about $0.5 {\mu}m$ for the MED and LOW cases,
indicating that some of the droplets in the grids where cloud water is larger 
than $10^{-3} g/kg$ did not exceed 
activation radius. the location of the parcels with large aerosol inside (but not exceeding 
activation radius) show that these parcels reside not only at the edge of the cloud, 
but also inside the cloud
and span the full spectrum of droplet sizes - from 1 to more than 100 ${\mu}m$. 
Having these parcels near the 
edge of the cloud is not surprising. Since each droplet sees the supersaturation at its location within the grid, 
at the edge of the cloud/clear air some of the droplets can evaporate and the radius can then drop below 
activation radius. Their presents inside the cloud  
may indicate an aerosol competing effect. 
Although for the big aerosol the activation supersaturation is small, the activation radius
is big and as a result more time is needed for these particles to grow to the activation size
(see \citeauthor{grabowski2011}, \citeyear{grabowski2011}). 
Some of the droplets 
on big aerosol can not even grow by condensation to reach the activation radius, since for aerosol
of 1 ${\mu}m$, the activation radius is about 35 ${\mu}m$. 

\begin{figure}[htb]
\begin{center}
\epsfig{figure=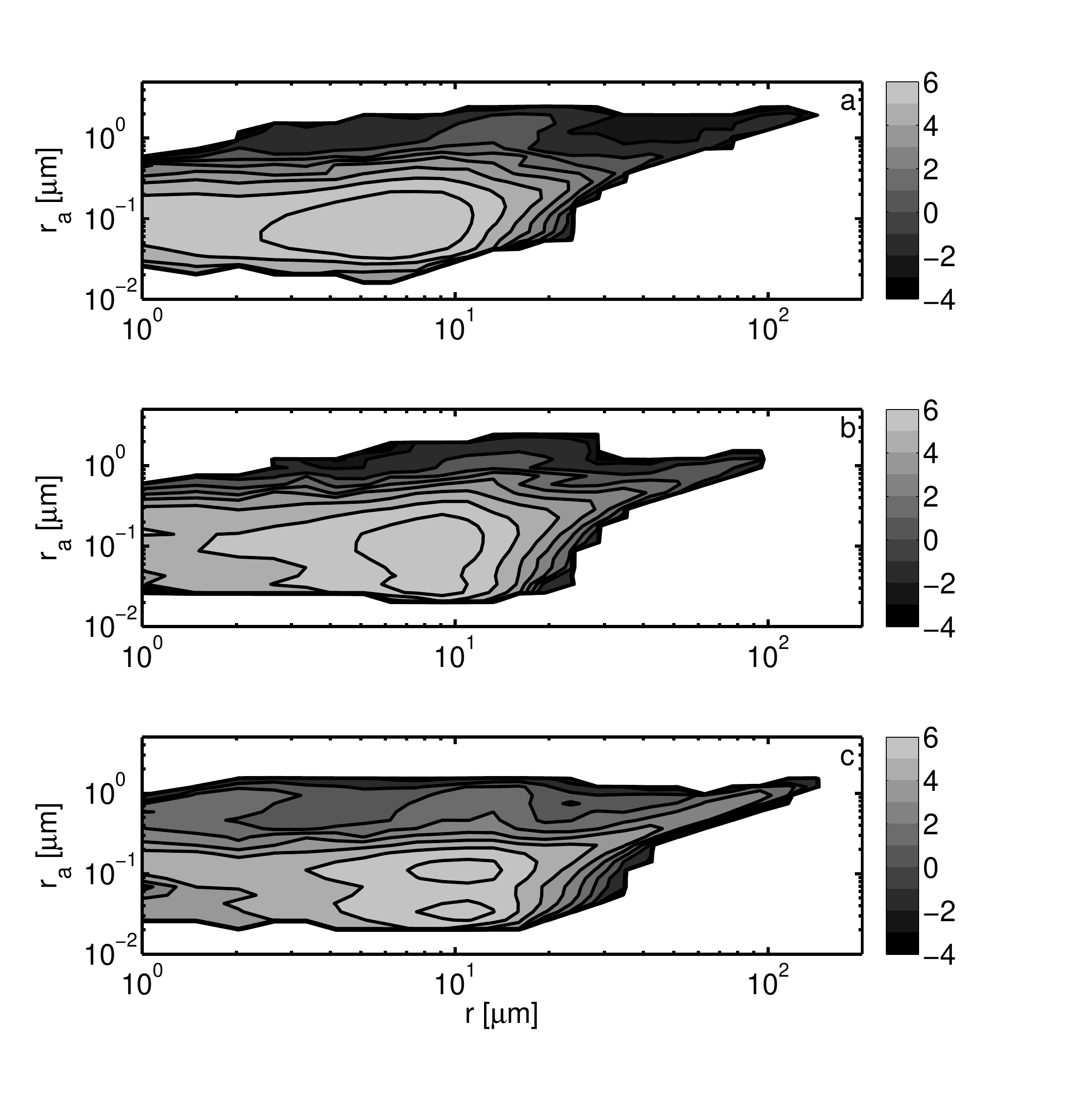,height=7in,angle=0}
\end{center}
\caption{  
$Log_{10}(N_d)$ as a function of aerosol and cloud droplet radius, for run HIGH - panel a,
run MED - panel b, and run LOW - panel c.
}
\label{aerosoldroplet}
\end{figure}

\noindent {\bf 4.2.3 Relation between aerosol and cloud droplet sizes} \\
In the model, full information about the aerosol and cloud droplet size is available. This data is presented on 
figure \ref{aerosoldroplet}, where $log_{10}$ of the droplet concentration is mapped on the collision grid for 
HIGH - panel a, MED - panel b and LOW - panel c cases. 
When integrated over the aerosol/droplet sizes the 2D cloud droplet concentration gives a cloud droplet 
spectrum or aerosol spectrum already discussed earlier in this paragraph. Figure \ref{aerosoldroplet}
shows that the majority of the droplets are on aerosol particles 
with sizes .02 - .3 ${\mu}m$. A significant change is present for aerosol sizes about 0.01 ${\mu}m$ 
with no aerosol particles having 
a significant amount of water below this size. Since the aerosol particles with sizes below 0.01 ${\mu}m$ activate 
very quickly (the activation radius is very small), it indicates that the maximum supersaturation in the simulations does 
not exceed the critical supersaturation for this size. Indeed the maximum supersaturation, of the order of  1\%, 
is observed 
at the end of the simulations, with a critical supersaturation for an aerosol radius 0.01 ${\mu}m$ being more than 2\%, at 283 K. 
The majority of the droplets 
on aerosol 0.02 - 0.3 ${\mu}m$ have a size of less than 20 ${\mu}m$, indicating that condensation is the main process responsible 
for the droplet growth for these aerosol sizes. Aerosol particles with sizes larger than 0.3 ${\mu}m$ are mainly inside 
the droplets with radius larger than 20 ${\mu}m$ and these droplets are created in a collision process. For aerosol 
radii bigger than 0.3 ${\mu}m$, with increasing
aerosol size the cloud droplet size also increases
since more collision events are needed to create big droplets. 

\noindent {\bf 5 Summary and Conclusions} \\
A Large Eddy Simulation model with a Lagrangian representation of microphysics 
is used to investigate the interactions between
cloud, aerosol, dynamics and thermodynamics for a cloud topped boundary layer. The microphysics in Lagrangian formulations
allow investigation of the cloud response to different aerosol concentrations and distributions, and provides 
detail information about aerosol-cloud droplet interactions. A distinct feature of this model is the treatment of
the coalescence process on a 2 dimensional grid. As a result not only droplet sizes, but also aerosol sizes 
are processed in the coalescence process. 

The numerical model captures general properties of the clouds, such as 
cloud droplet concentration and the profiles of cloud water, potential temperature and water vapour mixing ratio
observed during the VOCALS field campaign; showing that it is possible to model the  
relation between the aerosol concentration and distribution and cloud droplet number for a stratocumulus cloud. 
For other features, such as the droplet spectrum (both averaged inside the cloud
and on model levels), and the standard 
deviation of the cloud droplet distribution a difference between the observations and the model results are observed.
The numerical model under-predicts the standard deviation of the cloud droplet distribution  
for run with low cloud droplet concentrations compared to observations (1.5 $\mu$m vs. 3.3 $\mu$m). 
The inability of the model to produce a bigger standard deviation of the 
cloud droplet distribution for low cloud droplet concentration run may be result of the 2D set-up, 
since \citeauthor{andrejczuk2010},~(\citeyear{andrejczuk2010}) 3D simulation 
for very low aerosol concentration produced standard deviation close to 4 ${\mu}m$. 

The model results show that decreasing the cloud droplet/aerosol concentration doesn't lead to a significant
increase in either the largest cloud droplet sizes or the largest aerosol sizes. Since typically the biggest droplets 
have a big aerosol inside, one of the possible explanations is the removal of the big aerosol/droplets by drizzle.

This paper also shows the potential of the Lagrangian approach to microphysics to investigate cloud-aerosol interactions
and provides examples of the information possible to be derived from such a model; for instance the size of the cloud droplets
as a function of the aerosol size or the percentage of activated aerosol for different aerosol sizes. This information
may be used in the future to help interpret aircraft observations or quantify the relation between the aerosol concentration
and distribution and cloud droplet concentration and distribution. 
For each Lagrangian parcel the LCM model can distinguish 
between cloud droplets (having radius bigger than activation radius for given aerosol size) 
and aerosol with water (with radius smaller then activation radius). Model results show, that significant 
percentage of the droplets containing aerosol bigger than $0.3$ does not exceed activation radius. 
As a result these can behave different than activated aerosol (with radius bigger than activation radius) 
and for instance with the decreasing supersaturation evaporate. 

\begin{figure}[htb]
\begin{center}
\epsfig{figure=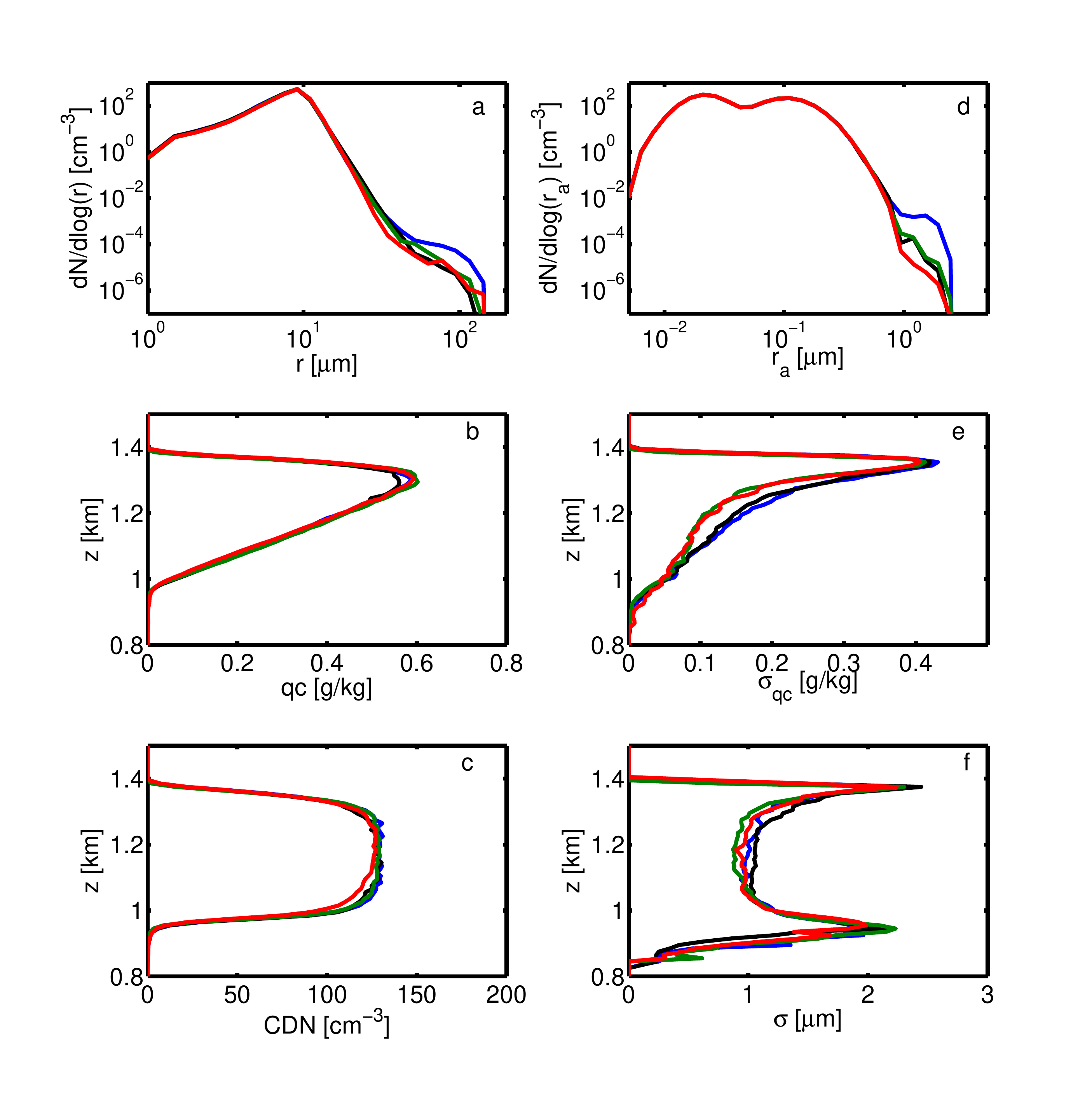,height=7in,angle=0}
\end{center}
\caption{
Sensitivity of the model solution for MED aerosol distribution 
to the number of parcels used to represent
initial aerosol distribution: 50 - blue, 100 - black, 200 - green, 400 - red. 
Panel a - droplet spectrum, panel b - cloud water mixing ratio profile, 
panel c - cloud droplet number profile, panel d - aerosol spectrum, 
panel d - profile of the standard deviation of the cloud water mixing ratio,
panel e - profile of the standard deviation of the cloud droplet spectrum.
All profiles are averaged over the last 3 hours.
}
\label{nfig12}
\end{figure}

\newpage
\noindent {\bf Appendix A: Sensitivity study}
As discussed and investigated in appendix of \citeauthor{andrejczuk2010}~(\citeyear{andrejczuk2010}) 
the Lagrangina microphysics depends on a set of parameters. One of these parameters is the number of Lagrangian
parcels used to represent the aerosol distribution in each model grid. In the simulations discussed 
here, 100 per model grid were used. 
For the MED case, sensitivity runs were performed to investigate how the model solution 
depends on number of parcels. As well as 100 per grid,
50, 200 and 400 Lagrangian parcels were placed in each model grid. Figure \ref{nfig12} shows values averaged 
over the last 3 hours of bulk properties of the simulated cloud. There is variability in the model solution 
on changing the number of parcels, but there is no consistent trends. Figure \ref{nfig12}a and \ref{nfig12}d, the blue curve, 
indicates that using 50 parcels per model grid may not be enough, because both aerosol and droplet processing in 
the collision procedure leads to higher concentrations of the large aerosol and cloud droplets than for the other cases.
The decrease of the standard deviation of the cloud water mixing ratio (${\sigma}_{q_c}$) 
with the number of parcels used to represent initial aerosol 
distribution is small. This indicates that the variability shown in figure \ref{qc_contour} is mainly 
due to dynamics. 
These are the only cases where consistent differences for runs with different number of parcels are observed. 
For other bulk properties, the differences are either small or show no trends indicating
that the variability in bulk cloud properties shown in figure \ref{nfig12} is a result of the different 
referred of the solutions resulting from the different forcing and interactions
between microphysics and dynamics/thermodymanics. 

\begin{figure}[htb]
\begin{center}
\epsfig{figure=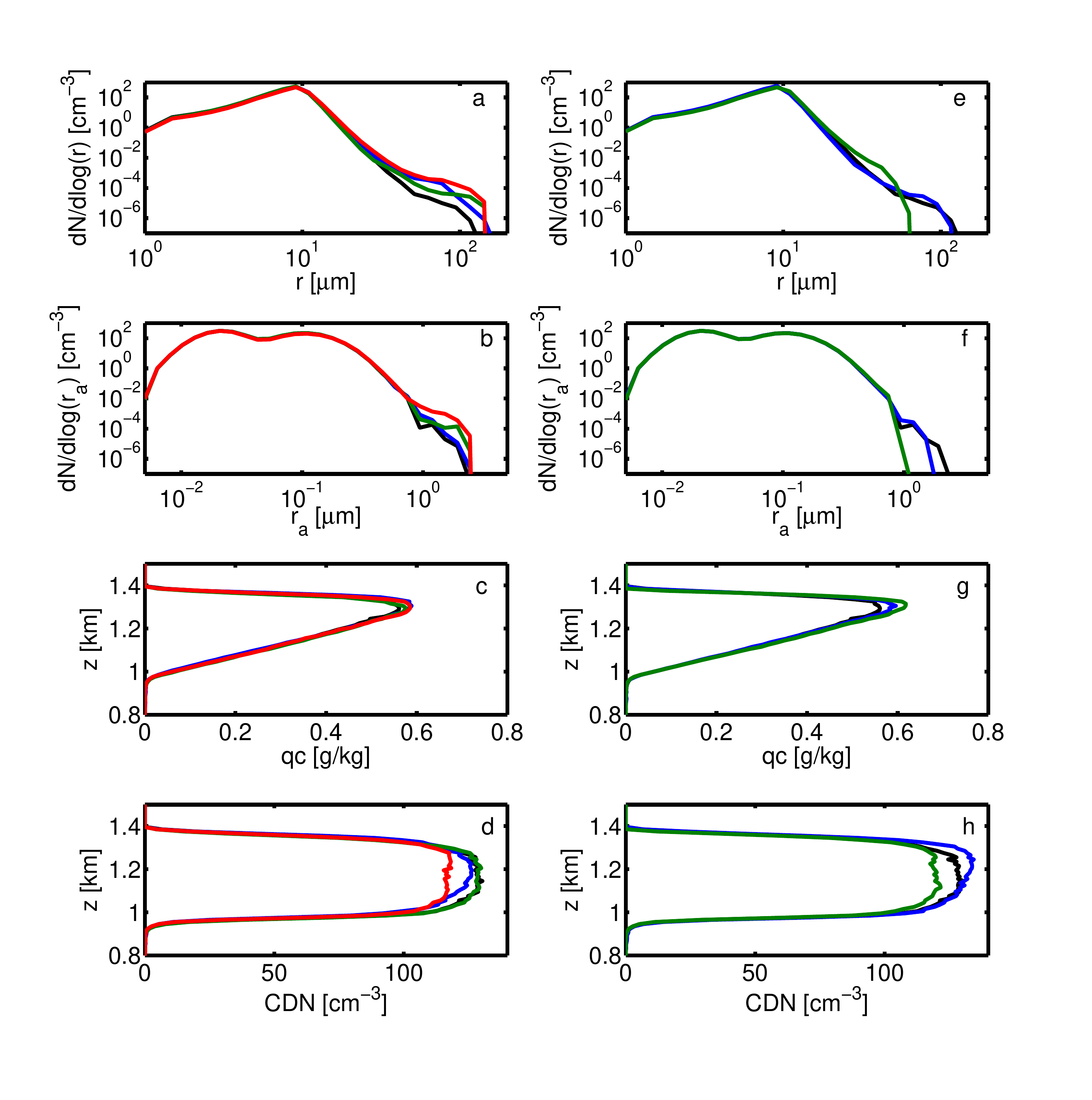,height=7in,angle=0}
\end{center}
\caption{
Left panels - mean profiles for the simulations: MED - black ($T_l$=4000, computational grid split into 4 
collision grids, parcels merged when difference in size is less than half of the bin width), 
simulation with $T_l$ = 1 - blue, simulation with the collision grid the same as computational grid - red,
simulation when parcels are merged for the difference in size less than  a quarter of the bin width - green.
Right panels - mean profiles for the simulations: MED - black (collision called every time step), 
collision called every 5 time steps (1s) - blue, collision called every 150 time steps (30s) - green.
All profiles are averaged over the last 3 hours. Panels a/e droplet spectrum, panels b/f - aerosol spectrum, 
panels c/g cloud water mixing ration profile, panels d/h - cloud droplet number profiles.
}
\label{nfig13}
\end{figure}

The other parameters solution depend are:
a threshold level (T$_l$) defining minimum number of cloud droplets
that the Lagrangian parcel represents, whether model grid is split into collision grids and how many parcels are
allowed in one bin before parcels are merged. Sensitivity of the solution to these parameters
is shown in figure \ref{nfig13}a-\ref{nfig13}d.
Decreasing T$_l$, not splitting model grid into 4 collision grids, and merging parcels when
the difference in size is less than the quarter of the bin size (half of the bin size in reference MED run) 
has a similar effect and leads to bigger drizzle sizes, higher 
drizzle concentrations and a larger concentration of the aerosol created in coalescence process
compared to MED reference run. 
Despite the differences in the cloud droplet and
aerosol distributions cloud water and cloud droplet number profiles are very similar.

Panels e-h of the figure \ref{nfig13} show the sensitivity of the solution to the frequency of the calls to collision procedure.
Beside the reference run, where the call is every time step two additional runs were performed, one with the 
call every 5 time-steps (1~s) and one with the call every 150 time-steps (300~s). 
Solution with the calls every 1s is very similar to the
reference solution, and the difference between these two solutions is within the variability for the solutions
discussed already. Solution with the call to the collision routine every 300~s produced much smaller drizzle sizes, but with 
higher than the reference (MED) drizzle concentrations and also processed almost no aerosol.
Additionally there is almost no speed-up (1\%) of the model execution when the collision procedure was
called every 300~s, compared to 15\% speed-up for the run with calls every 1~s. This indicates that the additional
part of the collision algorithm responsible for keeping number of physical droplets in each Lagrangian parcel
above threshold level has a significant effect on execution time. It follows that  
rather moderate time between calls to collision routine may be used to obtain faster model execution.      

\begin{figure}[htb]
\begin{center}
\epsfig{figure=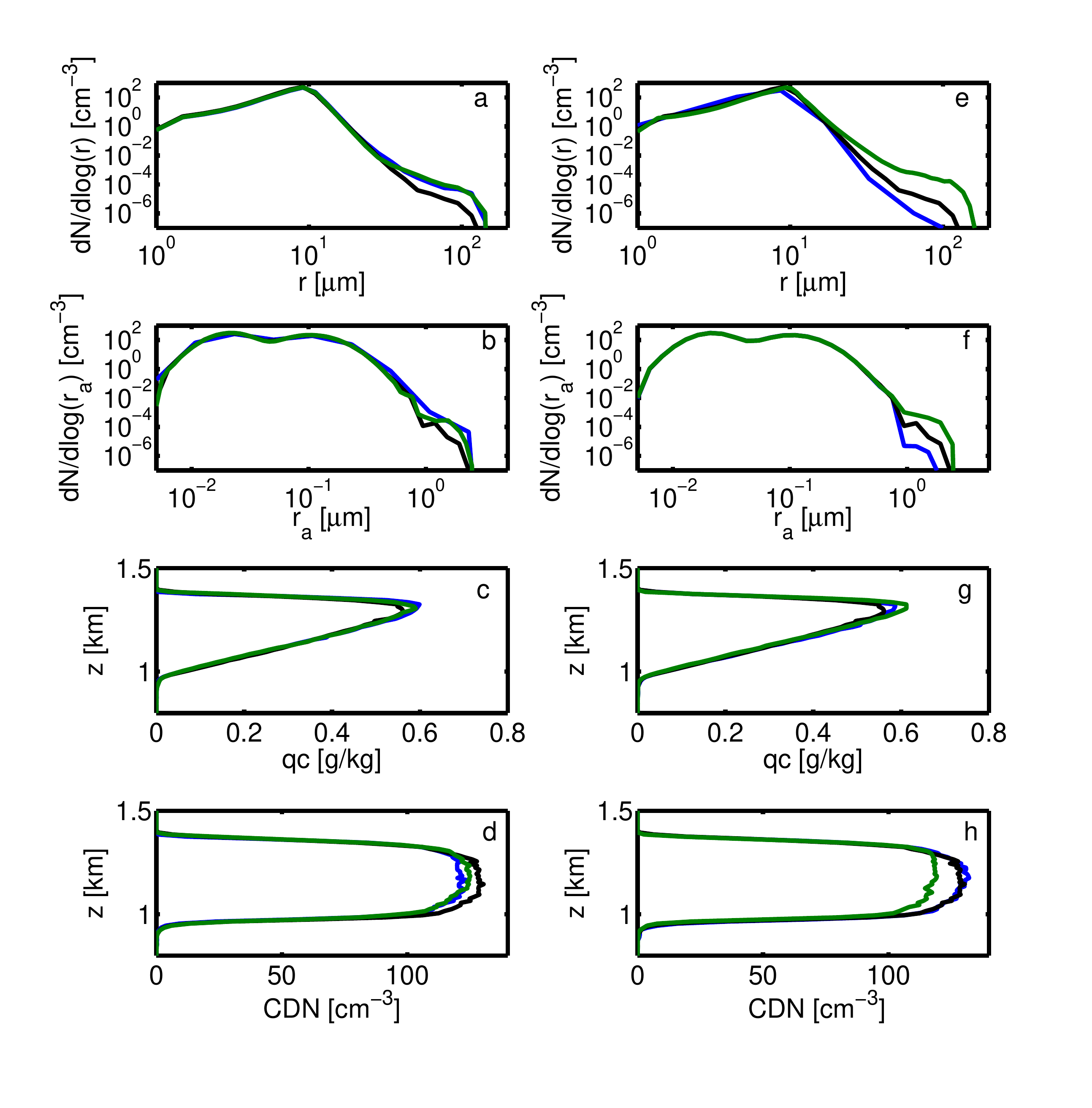,height=7in,angle=0}
\end{center}
\caption{
Left panels - mean profiles for the simulations with 30 droplet radius bins and 10 - blue, 30 (MED run) - black, and
60 - green, aerosol bins. Right panel - mean profiles for the simulations with 30 aerosol radius bins and
10 - blue, 30 (MED run) - black, and 60 - green, droplet radius bins. All profiles are averaged over the last 3 hours.
Panels a/e droplet spectrum, panels b/f - aerosol spectrum,
panels c/g cloud water mixing ration profile, panels d/h - cloud droplet number profiles.
}
\label{nfig14}
\end{figure}

In the LCM the droplet growth is grid free, but the microphysical grid is required to map collisions. As a result 
the solution depends on the number of bins used to represent the aerosol and the cloud droplet radius space. Results of the 
simulations shown in figure \ref{nfig14}a - \ref{nfig14}d show little sensitivity of the solution with different number
of bins (10, 30 and 60) used to represent aerosol sizes, when number of bins used to represent droplet spectrum is 30. 
The model is much more sensitive to the number of bins used to represent droplet spectrum for 30 aerosol bins
(\ref{nfig14}e - \ref{nfig14}h); 
where with a increasing
number of bins, the concentration and biggest drizzle sizes increases. The model also processes the aerosol 
more efficiently with an increasing number of droplet bins, 
as indicated by the increase of aerosol concentration for sizes bigger than 1 $\mu$m. The 
changes in the mean $q_c$ profiles are small, but use of the 60 bins to represent droplet sizes decreases cloud 
droplet number concentration. 
Although increasing number of bins tends to improve the solution 
(simulations using 30 bins under-predict drizzle concentrations compared to observations) it comes with an
increase of CPU time and model execution with 60 bins used to represent droplet sizes is almost two times more
expensive (increase by 90\%) than the run with 30 bins.

\begin{acknowledgement}
This work was supported by NERC grant NE/F018673/1. Observational data was provided by the 
British Atmospheric Data Centre (BADC). 
The authors would like to thank Dr Ian Crawford, Dr Jonathan Crosier and Dr Grant Allen 
from Manchester University 
for providing aerosol and microphysical data. 
The authors thank dr. Wojciech Grabowski for his comments.
This work made use of the facilities of HECToR, the UK's national high-performance computing service, which is provided by UoE HPCx Ltd at the University of Edinburgh, Cray Inc and NAG Ltd, and funded by the Office of Science and Technology through EPSRC's High End Computing Programme.
\end{acknowledgement}

\newpage

\begin{table}[htb]
\caption{Constants used to define profiles of the potential temperature, water vapour mixing ratio and cloud water mixing ratio.
}
\label{table.profiles}
\begin{center}
\tabcolsep .25in
\tabcolsep .075in
\begin{tabular}{lllllllllll}
\hline
\hline
$RUN$ & $z_B$ & $z_T$ & ${\theta}_B$ & ${\theta}_C$ & ${\theta}_T$ & $q_{vB}$  & $q_{vT}$  & $\alpha$ &  Observation & Date \\
      &  [m]  & [m]   &   [K] &  [K]  & [K]   &  [g/kg]   &  [g/kg]   & [K/m]    &  time [UTC]  & \\
\hline
\hline
HIGH & 800  & 1380& 291.1 & 293.0 & 302.5 & 8.3 & 0.3 & 3.3 $\times$ 10$^{-3}$& 11.08-11.28& 13 Nov. 2008\\
MED  & 1000 & 1400& 289.2 & 290.4 & 299.0 & 7.0 & 0.7 & 3.0 $\times$ 10$^{-3}$& 11.36-12.03& 31 Oct. 2008\\
LOW  & 900  & 1260& 290.1 & 291.1 & 301.0 & 7.8 & 0.7 & 2.7 $\times$ 10$^{-3}$& 11.47-12.02& 13 Nov. 2008\\
\end{tabular}
\end{center}
\end{table}

\begin{table}[htb]
\caption{Parameters of log-normal distributions observed/used in simulations. 
${\sigma}_{+}$, ${\sigma}_{-}$ are fits for the observed $mean$+/-${\sigma}$. 
LOW specifies aerosol distribution used in simulations (averaged for 
only 2 measurements). LOWA is the mean for all below the cloud observations 
with the corresponding fits for observations +/-${\sigma}$ given by
${\sigma}_{+}$/${\sigma}_{-}$.   
}
\label{table.lognormal}
\begin{center}
\tabcolsep .25in
\tabcolsep .075in
\begin{tabular}{lcllllll}
\hline
\hline
$RUN$ & & $N_1 [cm^{-3}]$& $r_1 [{\mu}m]$ & ${\sigma}_1$ & $N_2 [cm^{-3}]$ & $r_2 [{\mu}m]$&${\sigma}_2$ \\
\hline
\hline
HIGH&              & 380 & 0.071 & 0.45 & 160 & 0.029 & 0.31 \\
    &  ${\sigma}_{+}$  & 448 & 0.074 & 0.42 & 227 & 0.031 & 0.33 \\
    &  ${\sigma}_{-}$  & 327 & 0.064 & 0.53 & 78  & 0.027 & 0.27 \\
\hline
MED &              & 118 & 0.10 & 0.43 & 129 & 0.022 & 0.36 \\
    &  ${\sigma}_{+}$  & 212 & 0.10 & 0.56 & 151 & 0.022 & 0.31 \\
    &  ${\sigma}_{-}$  & 47 & 0.10 & 0.28 & 89  & 0.022 & 0.39 \\
\hline
LOW &              &  42 & 0.11 & 0.25 & 111 & 0.023 & 0.47 \\
\hline
LOWA&                 & 63 & 0.10  & 0.27  & 134  & 0.025   & 0.50    \\
    &  ${\sigma}_{+}$ & 84 & 0.10  & 0.28  & 170  & 0.025   & 0.54     \\
    &  ${\sigma}_{-}$ & 41 & 0.10  & 0.24  & 99   & 0.025   & 0.45     \\

\end{tabular}
\end{center}
\end{table}

\end{document}